\documentclass[%
 aip,
 amsmath,amssymb,
 reprint,%
]{revtex4-1}

\usepackage{physics}
\usepackage{graphicx}% Include figure files
\usepackage{dcolumn}% Align table columns on decimal point
\usepackage{bm}% bold math
\usepackage[utf8]{inputenc}
\usepackage[T1]{fontenc}
\usepackage{mathptmx}
\usepackage{etoolbox}

\usepackage[english]{babel}

\makeatletter
\def\@email#1#2{%
 \endgroup
 \patchcmd{\titleblock@produce}
  {\frontmatter@RRAPformat}
  {\frontmatter@RRAPformat{\produce@RRAP{*#1\href{mailto:#2}{#2}}}\frontmatter@RRAPformat}
  {}{}
}%
\makeatother

\renewcommand{\selectlanguage}[1]{} 

\begin{document}

\preprint{AIP/123-QED}

\title{Optimizing loading of cold cesium atoms into a hollow-core fiber using machine learning}
% Force line breaks with \\
\author{P. Anderson}
 \affiliation{Institute for Quantum Computing, University of Waterloo, Waterloo, ON, Canada}
 \affiliation{Department of Physics, University of Waterloo, Waterloo, ON, Canada}%Lines break automatically or can be forced with \\
\author{S. Venuturumilli}%
 \affiliation{Institute for Quantum Computing, University of Waterloo, Waterloo, ON, Canada}
\affiliation{Department of Electrical and Computer Engineering, University of Waterloo, Waterloo, ON, Canada}%
\author{T. Yoon}%
 \affiliation{Institute for Quantum Computing, University of Waterloo, Waterloo, ON, Canada}
\author{M. Bajcsy}
 \affiliation{Institute for Quantum Computing, University of Waterloo, Waterloo, ON, Canada}
 \affiliation{Department of Physics, University of Waterloo, Waterloo, ON, Canada}
\affiliation{Department of Electrical and Computer Engineering, University of Waterloo, Waterloo, ON, Canada}%
 \homepage{}

\date{\today}% It is always \today, today,
             %  but any date may be explicitly specified

\begin{abstract}
Experimental multi-parameter optimization can enhance the interfacing of cold atoms with waveguides and cavities. Recent implementations of machine learning (ML) algorithms demonstrate the optimization of complex cold atom experimental sequences in a multi-dimensional parameter space. Here, we report on the use of ML to optimize loading of cold atoms into a hollow-core fiber. We use Gaussian process machine learning in M-LOOP, an open-source online machine learning interface, to perform this optimization. This is implemented by iteratively adjusting experimental parameters based on feedback from an atom-counting measurement of optical ``bleaching''. We test the effectiveness of ML, alongside a manual scan, to converge to optimal loading conditions. We survey multiple ML runs to automatically access appreciable atom-loading conditions. In conjunction with experimental design choices, ML-assisted optimization holds promise in the implementation and maintenance of complex cold atom experiments.  
\end{abstract}

\maketitle

\section{Introduction}

Recently, there have been concerted efforts by researchers in the cold atom community to leverage machine learning (ML) algorithms to optimize their experiments. This includes optimization of temperature and densities (number and phase-space) of atoms in free-space magneto-optical traps (MOT) using the six-beam architecture \cite{tranter2018multiparameter,reinschmidt2023reinforcement}, as well as grating-based MOTs \cite{seo2021maximized}. ML-assisted optimization of Bose-Einstein condensates (BECs) have also been reported \cite{milson2023high,vendeiro2022machine,xu2024bose}. Additionally, ML has also been  \cite{gupta2022machine} utilized to interface atoms with optical nanofibers. The results outlined in this paper further the effort of using ML in optimizing the interfacing of cold atoms  with a waveguide. Specifically, our aim is to optimize the loading of cold cesium atoms into a hollow-core photonic-bandgap fiber with a 7 $\mu$m diameter core core, with the goal of creating an atomic cloud of extremely high optical depth \cite{Blatt2014_OnedimensionalUltracold} to be used for photon storage \cite{Peters2020_QMemory_HCF} and wavelength conversion \cite {Telecom2025FrequencyConversion}, as well as fundamental studies of light-matter interactions \cite{SPIEAlMaruf2024, SPIEPasharavesh2025}. \\ 

%Loading cold atoms into the core of this fiber is a process that can be explained in a relatively simplistic manner, but it is a much more complex procedure in practice. 

Our experimental setup along with the general outline of the optimization procedure is shared in Fig. 1. The experimental setup is based on the system reported in Ref. \cite{Taehyun} with a few upgrades that includes imaging the atomic cloud from two perpendicular directions, as well as light being coupled into the lower end of a hollow-core photonic crytstal fiber (HCPCF) using a lithographically-defined mechanical fiber splicer \cite{MarufOnChip2017} and a custom-designed objective to improve light coupling into and collection from the upper end of the HCPCF. ML-based optimization is tested across both versions of the experimental system, i.e., with either a free-space-coupled or an on-chip-coupled HCPCF. \\

We first prepare a laser-cooled atomic cloud above a piece of hollow-core fiber. The free-falling atoms are then guided into the core of the hollow-core using a far-off-resonant trap. After setting of the optical fields in relation to the positioning of the fiber in our experiment, we have a choice of nine experimental parameters that can be electronically set. We would like to query the optimal configuration of these parameters to realize optimal loading of atoms. Such an optimization procedure can be applied to alternative designs of laser-cooling and loading stages of the experiment. Fast optimization can also assist in continued (long-term) operation of the setup where the system may need to be intermittently re-optimized following adjustments, perturbations or fluctuations in the environment \cite{tranter2018multiparameter}. \\

Atoms that successfully enter the fiber's core must start with a proper combination of position and velocity to be trapped into the fiber-core \cite{yoon2020simulating}. This initial condition is informed by the cooling and shut-off stages of the experiment. In our experiment this is specified by the intensity, detuning and duration of the laser fields, along with the strength and ramping of the magnetic fields. A non-trivial choice of these settings can result in increased loading of atoms into the fiber. ML algorithms in cold atom experiments were able to discover canonical techniques without a-priori knowledge\cite{vendeiro2022machine} and allow for the exploration of non-intuitive working conditions for experiments\cite{tranter2018multiparameter}. \\

Here, we demonstrate an implementation of a Gaussian process machine learner using M-LOOP (Machine-Learning Online Optimisation Package). We contrast its performance alongside a manual scan and empirically survey multiple runs to reach optimal conditions in the given space. We also observe ML-optimization to quickly access experimental settings leading to $\sim 4 \times 10^3$ atoms being loaded into the fiber.

% intensity, detuning and duration of the laser fields, along with ramping of the magnetic fields

\begin{figure*}[ht]
\includegraphics[width=1.93\columnwidth]{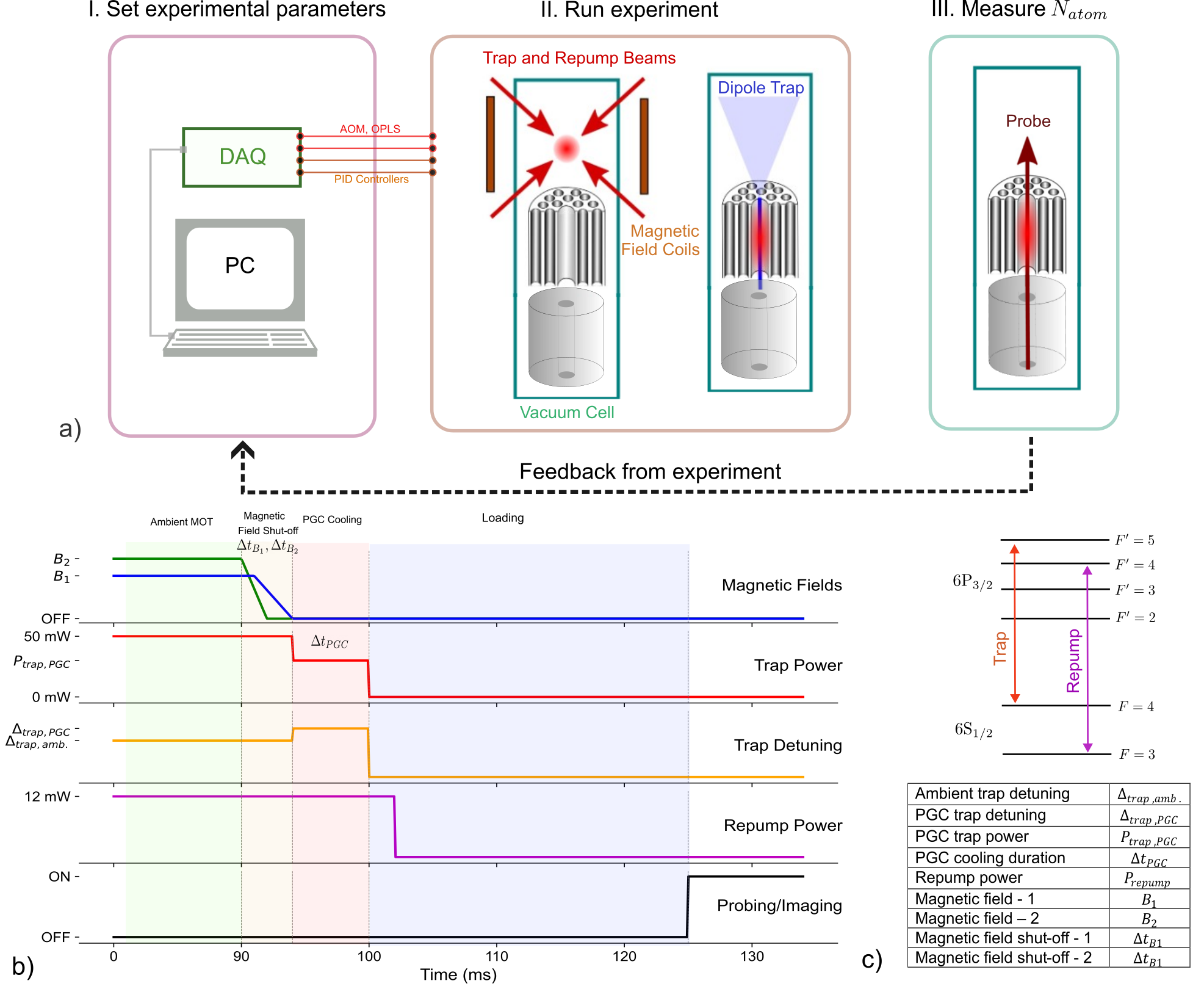}
\caption{a) Outline of the optimization: A PC + DAQ system communicates the digitally accessible parameters to the experiment. In the experiment, atoms are first laser-cooled and are then trapped into the HCPCF . Following which, the number of atoms inside the fiber ($N_{atom}$) are probed, which serves as a feedback to determine the next set of experimental parameters. b) Experimental sequence consisting of the laser-cooling fields' (trapping and repump) power and detuning, along with the strength and timing of the magnetic fields. c) Trapping and repump beam transitions in the Cs D$_2$ line and a tabulation of the parameters that we can currently access for optimization.}
\label{fig_setup}
\end{figure*}

%). These steps depend on parameters such as: the intensity and detuning of the lasers, magnetic field strengths, alongwith the duration of each of the steps (Fig. 2). Given the dependencies between several parameters, optimizing them may not be simple as scanning them individually.\\
%An intriguing solution to this otherwise complicated problem came from a couple of recent papers in cold atom physics, both of which use a software now known as M-LOOP. 

% \section{Experimental Methods}
% \begin{figure}[ht]
% \centering
% \includegraphics[width=\linewidth]{Setup.pdf}
% \caption{The experimental setup: atoms dispensed from a cesium source are first slowed and trapped, just above the fiber segment, by a magneto-optical trap. Typically 60 million atoms are trapped before being released via the turn off of trapping beams and magnetic field. Once released, polarization gradient cooling is employed to bring the cloud temperature to below 15 $\mu K$. These cold atoms will drop towards the core of a hollow-core fiber via gravity and a red-detuned attractive dipole beam. After loading, any necessary pump and probe beams  are coupled into the fiber and timed with the drop of the atoms. The probe is collected and sent to a SPCM after some polarization and spectral filtering.}
% \label{fig_setup}
% \end{figure}

\section{Experimental Methods}

A PC (LabVIEW) + DAQ (NI) system controls the experiment (Fig. 1) by setting parameters of the laser and magnetic fields. Laser field intensities and frequencies are set by acousto-optical modulators (AOM) and off-set phase-lock systems (Vescent), respectively. Magnetic fields are set via custom PID circuit boards. \\

A conventional six-beam magneto-optical trap setup is used to first prepare an ambient cloud of atoms in a vacuum system where the atoms are sourced by driving current through an alkali dispenser. Atoms then undergo sub-Doppler polarization gradient cooling (PGC) before being trapped into the HCPCF via a dipole trap. We use a 2-3 cm segment of HC-800B photonic-bandgap fiber produced by NKT Photonics as the HCPCF which enables us to guide light with wavelengths (750-950 nm) and atoms along its core (with diameter $\sim$7 $\mu$m). We employ polarization and spectral filtering to remove the dipole trap from probe measurements. The probe is then collected and sent to a photon counter (Excellitas SPCM-850) used with a time-tagging system (ID-900, IDQuantique) .

\subsection{M-LOOP: Online machine learning}

For our experiment, we utilize M-LOOP, a Python-implemented machine learning interface to optimize our experiment. The interface allows a user to choose from several optimization algorithms which, when fed experimental data during the experiment, iteratively attempts to optimize a measured output value based on any number of experimental inputs \cite{MLOOP1}. We consider Gaussian process (GP) regression across some of our parameter space and assess the speed of optimization. GP regression was shown to identify sensitive settings in a cold atom experiment optimizing BEC formation and also enabled a fast optimization which was performed within an hour \cite{MLOOP2}. GP regression can be beneficial in handling a relatively small amount of training data, contrasted with other methods like artificial neural networks \cite{MLOOP2}.

\subsection{Cost function: Atom number via `Optical bleaching'}
\begin{figure}[ht]
\centering
\includegraphics[width=\linewidth]{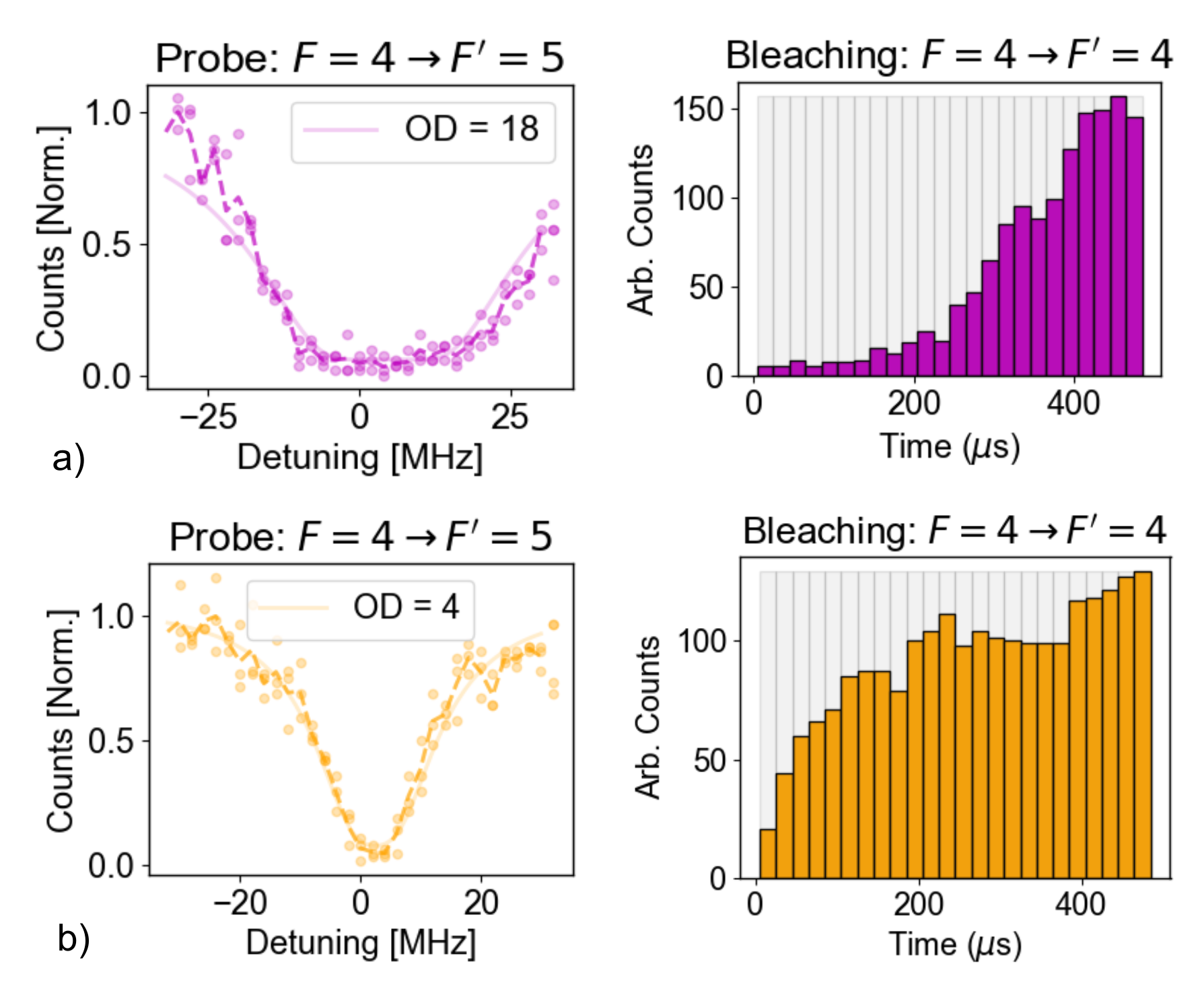}
\caption{a), b) Contrasting optical depth and bleaching measurements for two loading situations with estimated $N_{atom}$ = 1185 $\pm$ 279 (magenta) and 371 $\pm$ 105 (orange). (Left) Probing atoms along the cycling transition $\ket{F=4}\rightarrow\ket{F^\prime=5}$ to measure the loaded ensemble's optical depth. (Right) A bleaching measurement result when we send $\ket{F=4}\rightarrow\ket{F^\prime=4}$ light to the Cs atoms. Atoms absorb this light initially, and gradually scatter to the $\ket{F=3}$ state where the become transparent.}
\label{fig_cost_function}
\end{figure}

% \subsection{Optical Bleaching}
% \begin{figure}[ht]
% \centering
% \includegraphics[width=\linewidth]{Bleach_Hist_Levels.pdf}
% \caption{(left) A histogram of a train of resonant pulses travelling through the hollow-core fiber with atoms loaded (blue) and without (gray). The absence of counts in the blue histogram is due to atoms, loaded into the core of the fiber, scattering the photons. Counting the missing photons and taking into account the branching ratio will give the number of atoms loaded into the fiber's core. (right) Diagram for the hyperfine levels of interest for optical bleaching in cesium. By preparing our atoms in the $\ket{F=4}$ ground state and illuminating the $\ket{F=4}\rightarrow\ket{F^\prime=4}$ transition, each atom will scatter roughly two photons before moving to the $\ket{F=3}$ ground state. Once here, the atom will no longer interact with the probe light.}
% \label{fig_bleach_hist_levels}
% \end{figure}

We gauge the number of atoms ($N_{atom}$) inside the fiber using a single probe pulse sequence. If the probe sequence is long enough such that all the atoms are moved to a different ground state, the atomic ensemble becomes ``bleached" and transparent to the probe light\cite{Taehyun,blatt2014one,bajcsy2009efficient}. When we probe Cs atoms with light resonant to the $6S_{1/2} \ket{F=4}  \leftrightarrow 6P_{3/2} \ket{F^\prime =4}$ transition, atoms in the initially prepared $6S_{1/2} \ket{F=4}$ state can scatter to the $6S_{1/2} \ket{F=3}$ ground state with a branching ratio of 5/11, rendering the atom transperant to subsequent probe light. By examining the histogram of the pulse train after an attempt to load atoms, we can estimate the number of atoms by ``counting" the number of missing photons, given the collection efficiency of light from the HCPCF. Fig. \ref{fig_cost_function} contrasts two loading scenarios with frequency scan measurements across the cycling transition $6S_{1/2} \ket{F=4}  \leftrightarrow 6P_{3/2} \ket{F^\prime = 5}$ and bleaching measurements. The ML optimizer is set to minimize a cost-function which is defined either inversely ($\propto -N_{atom}$) or reciprocally ($1\propto 1/N_{atom}$). 

\section{Results}
\subsection{Gaussian process regression to optimize loading}

\begin{figure*}[ht]
\centering
\includegraphics[width=1.9\columnwidth]{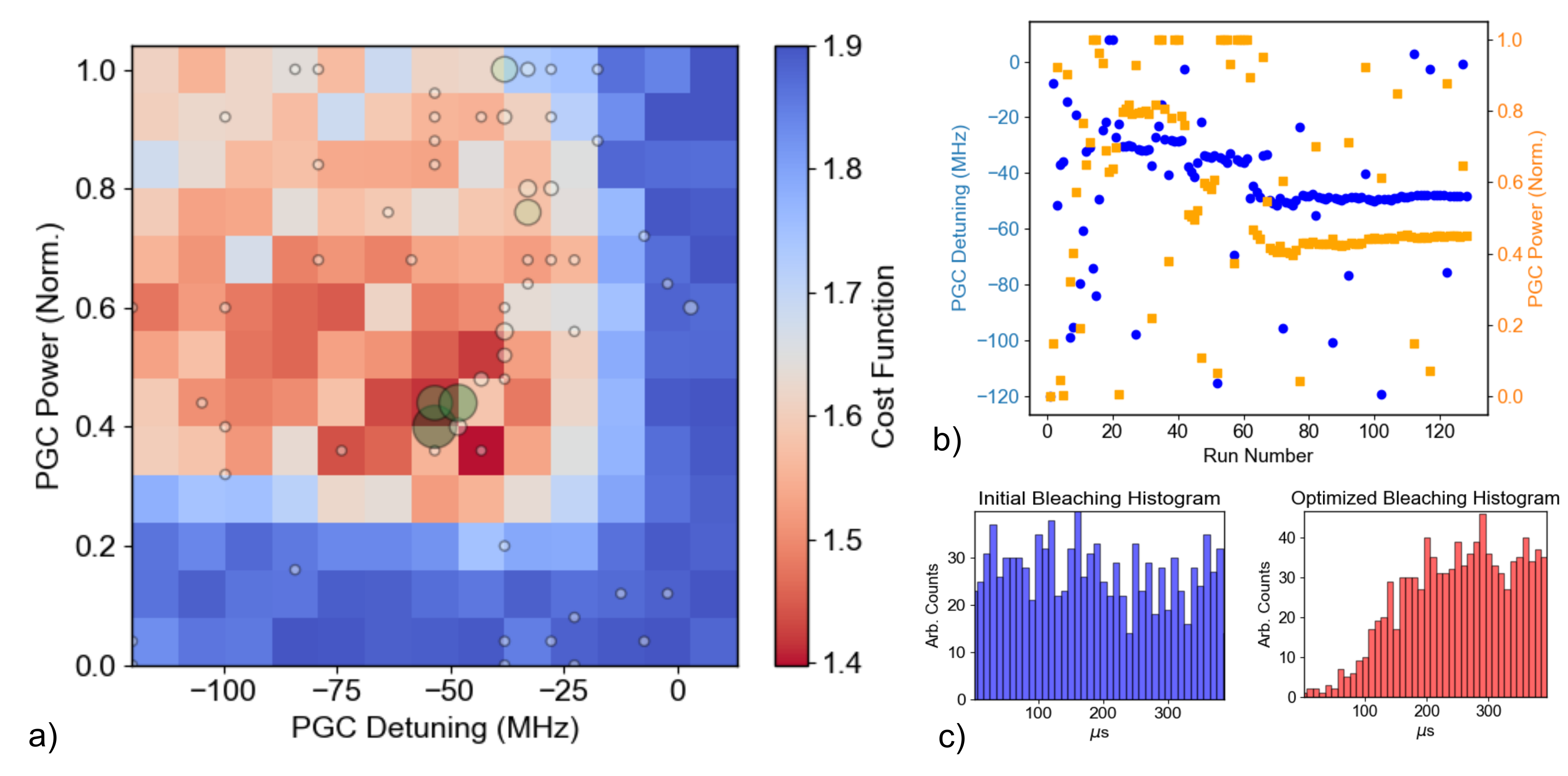}
\caption{a) Manual scan (blue-red) of loading of atoms versus intensity (power) and detuning of  the PGC stage of the experiment, overlayed (green circles) by the settings tested by the ML-optimizer during a single run. The size of the dot represents the number of times the algorithm queried the parameter. The colours correspond to a linear cost function, $2-\left(\frac{N_{atom}}{1000}\right)$, where $N_{atom}$ is the number of atoms loaded. b) Progression of the queried parameters across the run. c) Bleaching histograms collected at the beginning and at the optimal point of the run.}
\label{fig_2D_PGC_Manual}
\end{figure*}

We initially test an ML run to optimize loading into a free-space-coupled HCPCF alongside a manual two-parameter scan of the intensity (power) and detuning of the PGC stage of our experiment. The manual scan is performed by uniformly sampling and iterating over the two parameters, and is shown in Fig. \ref{fig_2D_PGC_Manual} a) (blue-red colour plot). One can see that the bottom and right edges of the plot show noticeably worse loading than the middle or top left. This can be inferred due to the limited cooling effect due to lower intensities (bottom) and heating due to resonant scattering when detunings are close to zero (right). The optimal choice over these two parameters is around  $\left(\Delta_{trap, PGC} = -54\text{ MHz},P_{trap, PGC} = 0.44 P_\text{cooling}\right)$, where $P_\text{cooling}$ is the power of the ambient trapping field $\approx$ 50 mW. The green overlayed circles follow an ML scan with the same parameters. Starting from virtually no atoms, we can converge to $N_{atom}$ $\sim$ 600 (Fig. \ref{fig_2D_PGC_Manual} c)). Fig. \ref{fig_2D_PGC_Manual} b) plots the evolution of the parameters over 128 runs where the central optimal region is probed on-goingly. \\

\begin{figure*}
\centering
\includegraphics[width=2\columnwidth]{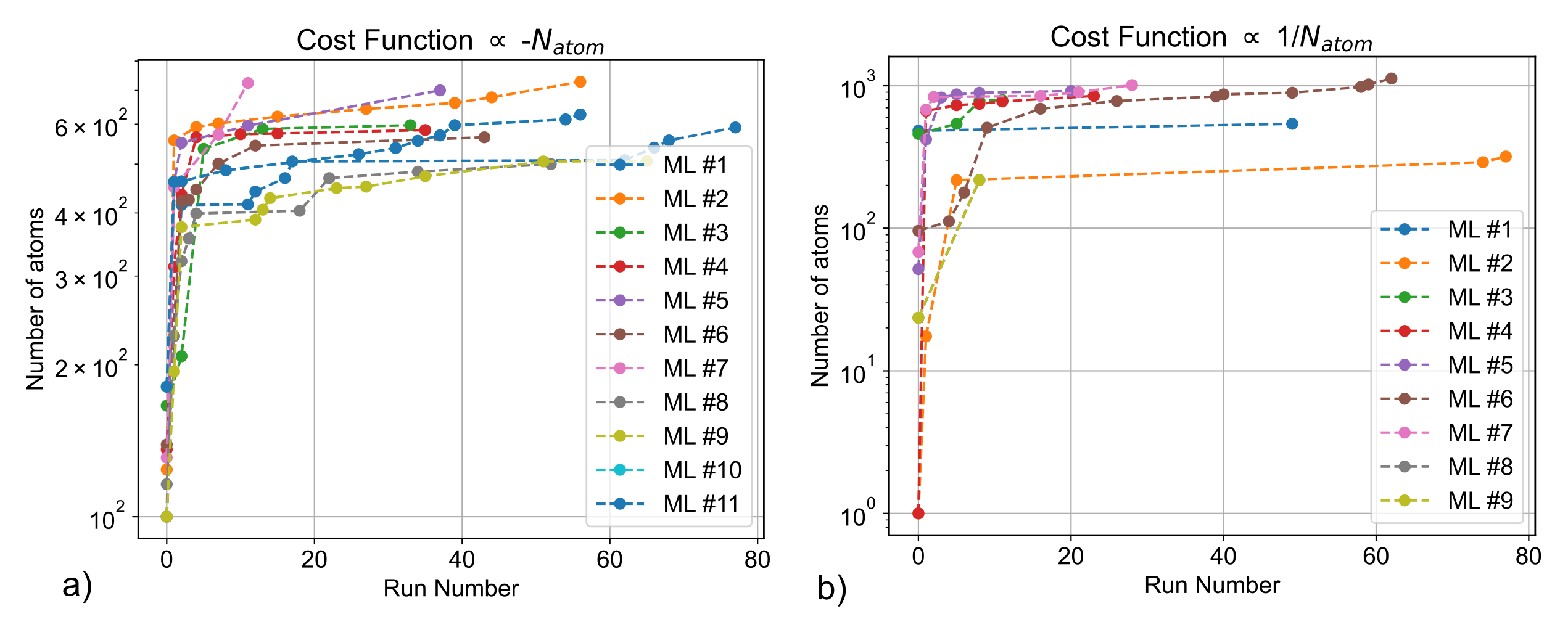}
\caption{Multiple ML runs for choices of a) linear and b) reciprocal cost functions to optimize loading via varying the cooling beam's intensity (power) and detuning.}
\label{fig_ML_cost}
\end{figure*}

Fig. \ref{fig_ML_cost} empirically surveys multiple ML runs to assess their convergence to comparable $N_{atom}$ using linear ($\propto -N_{atom}$) and reciprocal ($\propto 1/N_{atom}$) cost functions. The plots have been filtered such that only the next improvement in the cost function is plotted with the run number it occurred at. In between two non-consecutive runs will be cost function values that are worse or equal to the current best result in the particular experiment. In both cases, cost function values rapidly improve with diminished returns as more runs are performed. Several experiments from both sets of ML runs approach a similar loading quality of $N_{atom} \sim 600 - 700$. \\

\begin{figure*}
\centering
\includegraphics[width=2.1\columnwidth]{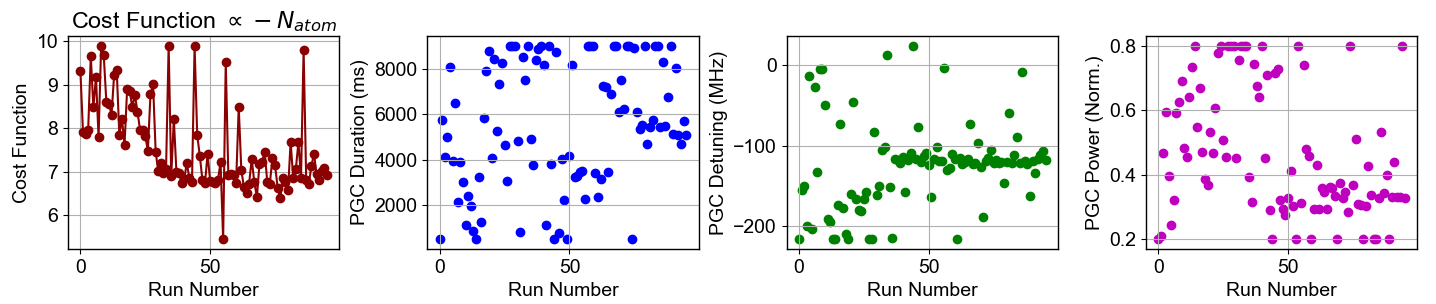}
\caption{Evolution of a single ML run where the intensity (power), detuning of the cooling laser and the duration of the PGC stage are the parameters being optimized for. (Left to right) Cost function, PGC duration, PGC detuning, PGC power (intensity) versus iteration (run) number. A linear cost function $10-\left(\frac{N_{atom}}{1000}\right)$ is chosen.}
\label{fig_ML_threeD}
\end{figure*}

Fig. \ref{fig_ML_threeD} illustrates a three-parameter evolution of an ML-optimization run for a distinct experimental system with an on-chip-coupled HCPCF when we also include the experimental parameter of the duration of the PGC cooling stage in the optimization process, leading to $N_{atom} \sim 4 \times 10^3$. 

\section{Conclusion}

We report a successful implementation of machine learning to optimize the loading of cold cesium atoms into a hollow-core fiber. Through a manual scan of a two-dimensional subset of our experimental parameter space, the accuracy of the M-LOOP software's algorithm was verified by its ability to find the same global minimum uncovered by the manual scan. A survey of multiple runs suggests convergence to good working conditions with no prior knowledge in reasonable time. We anticipate utilizing ML-based optimization over an extended parameter space, alongside alternative experimental laser-cooling and trapping sequences, to both explore non-trivial high atom-loading conditions and also, as a tool, to assist in maintaining system performance amidst adjustments, perturbations and environmental fluctuations.  

%This resulted in a 270\% increase in loading, the highest number of atoms loaded, with a 75\% increase in the number of runs compared to the 2-dimensional case.

\begin{acknowledgments}
This research was undertaken thanks in part to funding from the Canada First Research
Excellence Fund (CFREF) Transformative Qauntum Technologies  (TQT) initiative. Additionally, this work was supported by Industry Canada, NSERC Discovery grant, Canada Foundation for Innovation (CFI) Innovation Fund, Ontario Reseach Fund (ORF) Large Infrastructure grant, and by Ontario’s Ministry of Innovation Early Researcher Award.
\end{acknowledgments}

\section*{Data Availability Statement}

The data that support the findings of this study are available from the corresponding author upon reasonable request.

%\nocite{*}
\bibliography{optica_ML}% Produces the bibliography via BibTeX.

\end{document}